\documentclass[11pt]{article}
\usepackage{setspace}
\usepackage[utf8]{inputenc}
\usepackage{authblk}
\usepackage[hidelinks]{hyperref}
\usepackage{graphicx}
\usepackage{amsmath}

\begin{document}
	
	\title{Varying the explanatory span: scientific explanation for computer simulations
	} 
	
	
	\author{Juan M. Dur\'an\footnote{High Performance Computing Center Stuttgart - University of Stuttgart}\\\textit{International Studies in the Philosophy of Science} - (2017) 31(1):27--45\\ DOI:10.1080/02698595.2017.1370929.}
	
	
%
	\date{}

	\maketitle

	\begin{abstract}  
		\noindent This article aims to develop a new account of scientific explanation for computer simulations. To this end, two questions are answered: what is the explanatory relation for computer simulations? and what kind of epistemic gain should be expected? For several reasons tailored to the benefits and needs of computer simulations, these questions are better answered within the unificationist model of scientific explanation. Unlike previous efforts in the literature, I submit that the explanatory relation is between the simulation model and the results of the simulation. I also argue that our epistemic gain goes beyond the unificationist account, encompassing a practical dimension as well.
	\end{abstract}

\section{Introduction}

It is often claimed that computer simulations provide genuine instances of scientific understanding. These claims take note of the capacity of computer simulations to extend or enhance our access to the (empirical) world that is otherwise unreachable. Recent philosophical debates make use of notions like `explanation,' `prediction,' and `confirmation' to illustrate precisely how computer simulations ground such understanding. Unfortunately, such notions are typically embedded within larger philosophical discussions and are never addressed in detail. Take for instance scientific explanation, the chief notion of this article. In his discussion on how computer simulations produce new knowledge, Beisbart states: ``It is arguable that some scientific computer simulations provide explanations. If computer simulations are arguments and if explanations are arguments (or are at least built upon arguments), it is obvious how computer simulations can figure in explanation.'' \cite[429]{Beisbart2012}. El Skaf and Imbert \cite{ElSkaf2013} are another good example of how philosophers acknowledge explanation as an important epistemic feature of computer simulations, although they embed the notion in the discussion of thought experiments and scientific experimentation. To these authors ``[e]xperiments, computer simulations and thought experiments (hereafter E, CS and TE) are traditionally assigned different roles in scientific activity. For example, TE are often seen as ways of exploring conceptual apparatus and developing theorizing (Kuhn 1964), and CS as ways of providing theoretical explanations or making predictions, which E hardly contribute to.'' \cite[3452]{ElSkaf2013}. 

In this article, I propose to look squarely at the logic of scientific explanation for computer simulations. To this end, I expand the explanatory base of the unificationist approach and accommodate computer simulations within its framework. I then focus on answering two questions, namely: `what is the explanatory relation for computer simulations?' and `what kind of epistemic gain should we expect?'. Along with these questions, several philosophical issues relating to scientific explanation and computer simulations emerge and are discussed.

The discussion is organized as follows. Section \ref{SE_and_CS} discusses the literature on scientific explanation and computer simulations, as well as establishes the aims of this article. A chief result of this section is that my approach departs from earlier attempts in the literature, both in terms of the explanatory relation as well as in the treatment of the epistemic gain. Section \ref{Example} elaborates on an example of a computer simulation of an orbiting satellite under tidal stress. Admittedly, the example is rather simple, but it helps to introduce the basic terminology and makes the analysis of explanation a manageable task. The treatment of more complex computer simulations are discussed in section \ref{Conclusions}. Section \ref{Explaining_SP} addresses the logic of scientific explanation for computer simulations at face value. It begins by arguing in favor of the unificationist account of scientific explanation as a suitable theoretical framework for computer simulations. Section \ref{The_unificationist_framework} presents the unificationist account as elaborated by Kitcher (\cite{Kitcher1981a}, \cite{Kitcher1989}) and how computer simulations are accommodated within this framework. Section \ref{Explanatory_relation} fleshes out the form that an explanatory relation has for computer simulations using the example discussed in section \ref{Example}. Finally, section \ref{understanding} contains a discussion of the kind of epistemic gain obtained by explaining with computer simulation. 

\section{Computer simulations and scientific explanation} \label{SE_and_CS}

Recent efforts towards unveiling the logic of scientific explanation for computer simulations can be found in the work of \cite{Weirich2011} and \cite{Krohs2008}. Both authors defend a similar view on the issue. First, that the mathematical models implemented as the computer simulation have explanatory force (i.e., the mathematical model is identified as the \textit{explanans}). Second, that explanation is of a real-world phenomenon (i.e., real-world phenomenon is identified as the \textit{explanandum}). Thus understood, Weirich and Krohs subscribe to the standard view on scientific explanation rather than giving to computer simulations a role in the explanatory relation. One could then ask which is the explanatory role played by computer simulations? The answer is, for both authors, the same: a computer simulation plays the instrumental role of computing the set of solutions for the mathematical model. 

In the following, I discuss each author's position individually and show that both wrongly take the explanans to be a mathematical model and the explanandum to be a real-world phenomenon. I then argue that the explanans must be identified with the simulation model -- that is, the model at the basis of the computer simulation which takes from, but cannot be identified with, a mathematical model -- and that the explanandum must be identified with the results of the computer simulation.


Let us begin with Weirich, who believes that computer simulations \textit{exemplify} the dynamics of mathematical models that represent a phenomenon under investigation. That is to say, computer simulations implement and compute the structural properties of mathematical models in a rather straightforward way \cite[156]{Weirich2011}. In this respect, there is little difference between one and the other, except for the fact that computer simulations facilitate solving the set of equations in the mathematical models. Weirich illustrates this point in the following way: ``A simulation [...] uses a dynamic model of a natural system that produces the phenomenon. It uses the model to imitate the natural system's production of the phenomenon'' \cite[156]{Weirich2011}. Thus understood, Weirich has good motivations for identifying the dynamic model as the explanans since it holds the right structures that describe the real-world phenomenon. Furthermore, since computer simulations perform only the instrumental role of computing the dynamic model, it is difficult to argue for them having any explanatory force.


As for the explanandum, we need to look at the interpretation of `producing the phenomenon' in the previous quotation. Weirich understands this notion as referring to the set of results of the computer simulation that robustly represent real-world phenomena \cite[158]{Weirich2011}. This is an important point that underscores two issues and that I later use to ground my claim that explanation is actually of the results of the simulation: first, that our access to the world using computer simulations is mediated by the results of the simulation; and second, that such results need to be verified and validated in order to count as representing real-world phenomena. Unfortunately, Weirich takes these ideas too far, conflating robust results with the actual real-world phenomena. That is to say, the results of computer simulations are a conceptual `bridge' that connects to the real-world phenomena to be explained. In and by themselves, the results have little value other than as a link to the real-world. Thus understood, it is not surprising that the author prefers a familiar explanandum, that is, to explain the actual real-world phenomena. He makes this point plain in the following way: ``I examine only simulations that a model guides, in particular, simulations in which the guiding model aims to explain a natural phenomenon'' \cite[156]{Weirich2011}. 

As argued, Weirich has no motivations to modify the standard explanatory sche\-mata of a model explaining real-world phenomena. To his mind, the use of computer simulation is only justified for finding the set of solutions of the dynamic model, and the results are so robust that they altogether represent the real-world phenomena. Weirich condenses some of these ideas: ``[f]or the simulation to be explanatory, the model has to be explanatory'' \cite[Abstract]{Weirich2011}. The example chosen also illustrates his viewpoint: ``a computer simulation of an economic market explains the emergence of an efficient allocation of goods if the model it follows does'' \cite[156]{Weirich2011}. As I discuss later, my account significantly differs from Weirich both in the methodology and the epistemology of computer simulations, as well as the representational role of results of computer simulations. If I am correct, then computer simulations do have explanatory force and there is a way to show it.


Krohs shares with Weirich the general picture of scientific explanation for computer simulations. However, he has a different stand on the methodology and epistemology of computer simulations, as well as the way results represent. First, Krohs agrees with Weirich that computer simulations only provide numerical solutions to mathematical models that represent real-wold phenomena \cite[278]{Krohs2008}. In fact, he explicitly adopts Hartmann's viewpoint by stating that the primary element of interest in computer simulations is the theoretical model whose set of solutions are to be found \cite[278]{Krohs2008}.\footnote{Hartmann famously argued that computer simulations are methods for solving in the computer the set of the equations of dynamic models \cite[83]{Hartmann1996}.} However, he has a different viewpoint regarding the relation between theoretical models and computer simulations.\footnote{To Krohs, `theoretical models' are mathematical models that describe a given empirical target system \cite[278]{Krohs2008}.} To his mind, a theoretical model suffers a series of modifications, falsifications, and distortions before any implementation on the computer is possible. Such modifications, Krohs argues, prevent us from claiming that the theoretical model and the computer simulation share the same structures. In Krohs' own words, ``[...] the model that constitutes the basis of the simulation is not identical to the theoretical model. The simulation does not strictly rely on the mechanism described by the theoretical model. It refers to an ad hoc modified mechanism, which is not supposed to occur in the real-world system.''\footnote{Let it be noticed that Krohs refers to structures and mechanisms indistinguishably. This is so because he frames computer simulations within the mechanicistic theory \cite[282]{Krohs2008}.} \cite[282]{Krohs2008}. Thus understood, Krohs has good reasons to identify the explanans with the theoretical model, as well as to block the simulation model for playing any explanatory role. It is the theoretical model, and not the simulation model, that holds the right set of structures for explanatory purposes. Furthermore, the ad hoc modifications reflect the unreliable source of information that is the simulation model \cite[280]{Krohs2008}. To dispel all doubts about his position, Krohs says ``the simulation model does not provide an acceptable explanation of the material system'' \cite[282]{Krohs2008}. 

From here it follows that the results of the simulation must forcefully vary from those expected from the theoretical model, even if in small amounts, and thus can never describe the mechanisms in the real-world phenomenon \cite[281]{Krohs2008}. Krohs puts this idea in the following way: ``results [of a computer simulation] deviate from those to be expected from the theoretical model by the sum of discretization and numerical errors'' \cite[282]{Krohs2008}. At best, Krohs says, also following Hartmann, results of the simulation `imitate' within certain acceptable degree the behavior of a real-world phenomenon \cite[283]{Krohs2008}. Thus understood, results of computer simulations cannot be an appropriate constituent in an explanatory relation that tries to account for an empirical system. Rather, the real-world phenomenon itself must then be identified with the explanandum.


It is interesting to note that, unlike Weirich who virtually neglects the role of computer simulations in the explanatory relation, Krohs suggests he has found a place for them. He asserts that ``in the triangle of real-world process, theoretical model, and simulation, explanation of the real-world process by simulation involves a detour via the theoretical model'' \cite[284]{Krohs2008}. Now, such a claim seems to be at odds with Krohs' previous position on the (small) role of computer simulations in the explanatory relation. In fact, it begs the question of how such a detour is possible at all, and what does it mean for explanatory purposes. Unfortunately Krohs neither provides an answer to these issues nor makes explicit the explanatory role of computer simulations. Instead, computer simulations play an instrumental role (i.e., made visible in the notion of \textit{simulacra} \cite[285]{Krohs2008}), and explanation is a relation between a theoretical model and real-world phenomena.



To my mind, neither Weirich nor Krohs fully capture the explanatory power of computer simulations. As mentioned, to these authors explanation takes the standard format of a mathematical model explaining real-world phenomena. Strictly speaking, computer simulations do not play any role in the explanatory relation, but only in solving the mathematical model as a means to relate with real-world phenomena. Such a stand is possible because there is a fundamental misunderstanding on the epistemological role of computer simulations, as well as the way in which researchers access a world mediated by their results. In this respect, my proposal differs from Weirich and Krohs on two accounts. On the one hand, I take the simulation model -- and not an exogenous mathematical model -- to be the \textit{explanans} in the explanatory relation; on the other, I take the results of computer simulations to be the \textit{explanandum}. Let me now make these points more clear.


Much has been said about the role that computer simulations occupy in scientific research. Many philosophers have claimed that computer simulations implement a mathematical model \textit{simpliciter}, and thus their use is justified when analytic methods are unavailable (e.g., \cite{Hartmann1996}, \cite{Humphreys1990a}, \cite{Guala2002}, and \cite{Parker2009b}, among others). This viewpoint grants computer simulations only a secondary epistemological value and thus the theoretical model must be preferred when possible. Weirich quite straightforwardly adopts this viewpoint, and so does Krohs by the longer road of claiming for ad hoc modifications in the simulation model. 
 
A closer view on the practice and methodology of computer simulation shows otherwise. Standard computer simulations involve the transformation of a manifold of mathematical models for their implementation as a simulation model. Such transformations are not merely formal (e.g., requiring discretization methods), but involve a series of contrivances as well. A simple example is illustrated in Figure \ref{Fig:Satellite_three_masses}, where the mass of the satellite is no longer treated as one, but rather divided into three masses connected by springs. Techniques of simplifying assumptions, removing degrees of freedom, and even substituting simple empirical relationships for more complex ones is an ubiquitous practice across computer simulations that Winsberg calls `ad hoc modeling' \cite[282]{Winsberg1999a}. Similar claims can also be founded in the work of \cite{Humphreys2004}, \cite{Morrison2009}, \cite{Winsberg2010} and, more recently, \cite{Duran2017b}. Following these authors, I show in section \ref{Example} that none of these transformations entail downplaying the epistemological value of computer simulations.


%

Another important issue overlooked by both Weirich and Krohs is that the significance of computer simulations comes primarily from providing information about a target system that is not available beforehand. Such information comes in two different ways. Either results represent real-world phenomena, in which case our access to the latter comes via the former, or the results do not have an empirical counterpart in the world, in which case the results are, in fact, all the information available. A vivid example of the first case are simulations that scale down time and space such as computer simulations on the evolution of life on Earth. Our information about the evolution of a system is first -- and, in this case, only -- accessible via the results of a computer simulation. An example of the second case is a simulation where the gravitational constant is set to $G=2m^{3}kg^{-1}s^{-2}$. To the best of our current scientific knowledge, no physical systems exist with such a value for the gravitational constant. The results of the simulation, whatever they might be, are nomologically impossible since they violate a physical constant. 
Either way, scientific practice involved with computer simulation is concerned with the use of results as their primary -- and sometimes only -- source of information. 

A chief philosophical problem here stems from finding ways to epistemically ground the results of the simulation as reliably representing real-world phenomena. This is the object of much dispute among philosophers. Authors like \cite{Morgan2003, Morgan2005}, \cite{Morrison2009}, and \cite{Parker2009b} have this issue in the background of their discussions on the epistemological power of computer simulations (for a discussion on this issue, see \cite{Duran2013c}). And several authors have tried to ground representation by looking at verification and validation methods (e.g., \cite{Oberkampf2010}, \cite{Kueppers2005}, and \cite{Lenhard2007}). More recently \cite{Bueno2014} has expanded his inferential conception to include computer simulations as inferential devices that represent empirical phenomena. 

It is standard for philosophers to take representation as a matter of degrees and of encompassing the totality of the results of the simulation. While the former point is correct, the latter is not always so. The overlooked issue is that the results of computer simulations do not always \textit{altogether} represent what they show. To illustrate this point, consider the visualization of Figure \ref{Fig:Spikes}. On the one hand, the spikes represent, to some acceptable degree, the behavior that a real-world satellite would produce due to tidal stress. On the other, we see a steady downward trend that is the product of a roundoff error coded in the simulation model. If the entire set of results are taken to represent real-world phenomena, as Weirich and Krohs have, then we are wrongly ascribing a trend towards a circular orbit to the behavior of the real-world satellite, when in fact it is an artifact in the computation of the simulation model. 

These considerations give us a purchase for identifying the \textit{explanans} and the \textit{explanandum} in the explanatory relation. I take the explanandum to be the results of the computer simulation for two reasons. First, because the use of computer simulations entails that our information about the target system comes first -- and sometimes only -- through the results of the simulation. In fact, this is to many the main reason for using computer simulations in the first place. A logic of scientific explanation for computer simulations must reflect this fact. That is, that explanation is first of the results of the simulation and later of the real-world phenomenon. Second, because \textit{directly} explaining real-world phenomena, as Weirich and Krohs suggest, requires the results to represent altogether the real-world phenomena. As shown before, results of computer simulations carry with them artifacts coded in the simulation model that cannot be ascribed to real-world phenomena and thus are not constitutive for their explanation. Ignoring these issues not only begs the question of the role of computer simulation in the explanatory relation, but also of which phenomenon is actually being explained (i.e., Weirich and Krohs have lost the ability to distinguish numerical artifacts from correct representation). As I show in the remaining sections, there is a way to explain the spikes and the steady trend separately.

This last point has some kinship with the reasons for taking the explanans to be the simulation model. It is the simulation model -- and not an exogenous mathematical model -- the unit with the most explanatory relevance for the results of the simulation. Although mathematical models play a significant role at the early stages of designing and coding the simulation model, these include and exclude structures which are not present in the mathematical model but that have explanatory force. Take again Figure \ref{Fig:Spikes}. As I show by the end of section \ref{Example}, the mathematical model is unable to account for the spikes trending steadily downwards. Similarly, an explanation of the length of the spikes depends, \textit{inter alia}, on the length of the spring that make the representation of the satellite in the simulation model in Figure \ref{Fig:Satellite_three_masses} -- instead, the satellite in the mathematical model is represented by a point.

By interpreting the explanans and the explanandum in this way, I believe we can make full sense of the logic of scientific explanation for computer simulations. Following Weirich and Krohs, my project also aims to account for real-world phenomena. I deviate from them, however, in that I fully include computer simulations as a component in the explanatory relation. Let me now discuss in more detail the form that such an explanation would take. I begin by presenting a simple example.

\section{Background terminology and an example\label{Example}}

\begin{figure}
	\centering\includegraphics[scale=0.35]{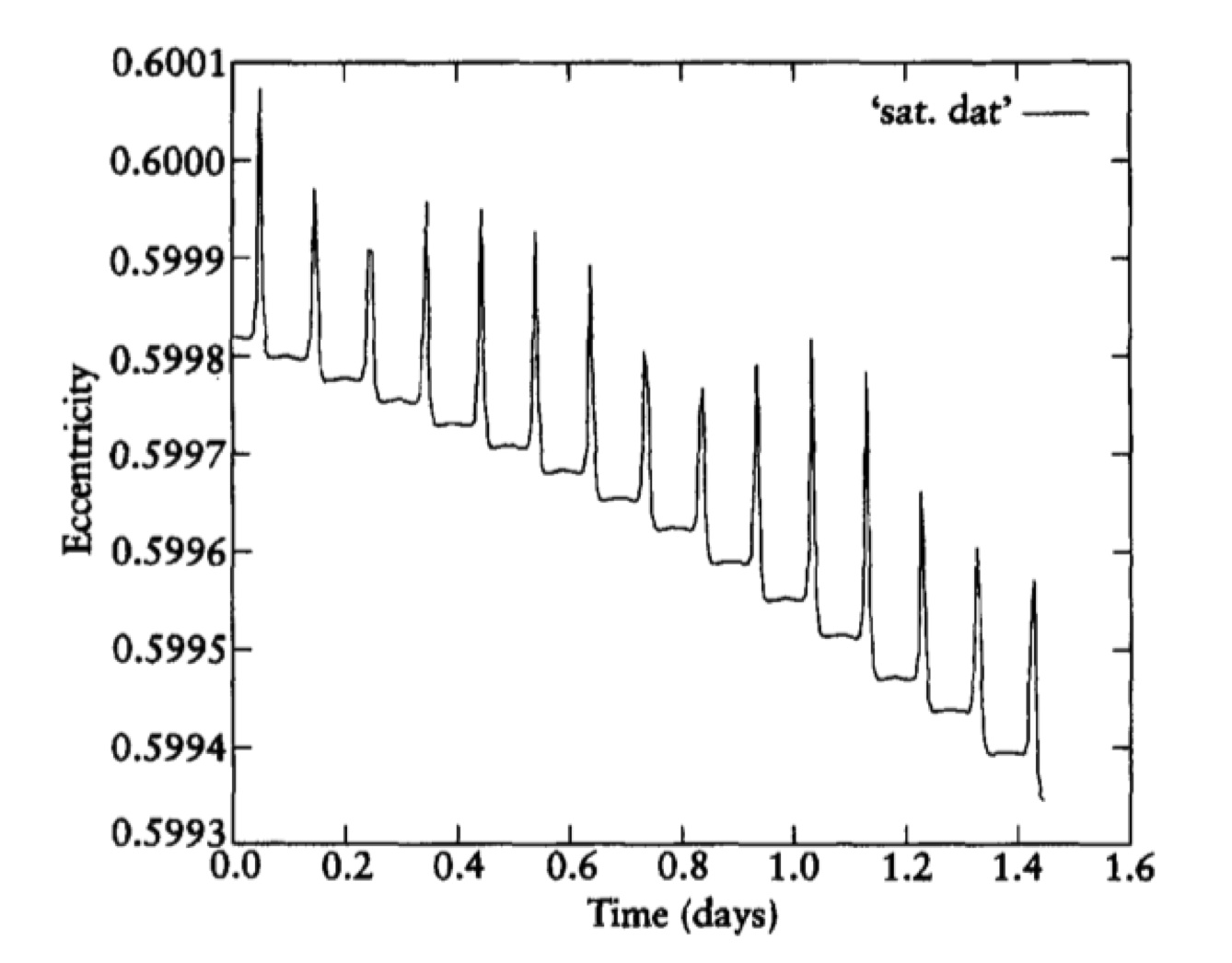}
	\caption{The orbital eccentricity as a function of time \cite[20]{Woolfson1999a}.}
	\label{Fig:Spikes}
\end{figure}

To introduce the base terminology, I make use of an example of a satellite orbiting around a planet. Although a simple simulation, it has considerable explanatory value. Following Woolfson and Pert \cite[17]{Woolfson1999a}, consider an orbiting satellite under tidal stress which stretches along the direction of the radius vector. This model presupposes, in addition, that the orbit is non-circular, and therefore, that the stress is variable making the satellite expand and contract along the radius vector in a periodic fashion. Since the satellite is not perfectly elastic, the mechanical energy is converted into heat, which is radiated away. The overall effect is, however, that although mechanical energy is lost, the system as a whole conserves angular momentum. The results of the simulation are shown in Figure \ref{Fig:Spikes}. The following conditions and equations are included in the model:\\

\begin{figure}
	\centering\includegraphics[scale=0.4]{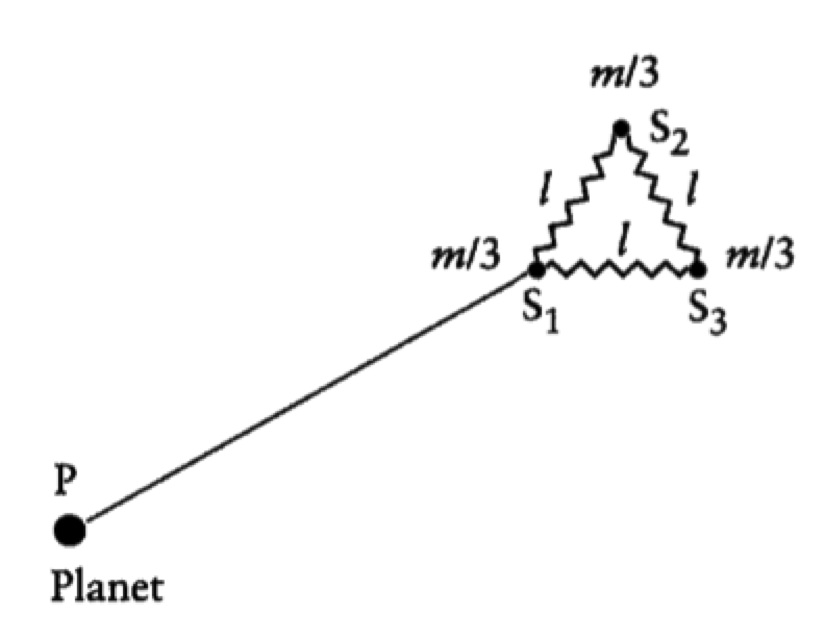}
	\caption{The satellite is represented by three masses, each $m/3$, connected by springs each of the same unstrained length, $l$ \cite[19]{Woolfson1999a}.}
	\label{Fig:Satellite_three_masses}
\end{figure}

\begin{quotation}
	\noindent For a planet of mass $M$ and a satellite of mass $m$ $(\ll M)$, in an orbit of semi-major axis $a$ and eccentricity $e$, the total energy is
	
	\begin{equation}
	E=-\frac{GMm}{2a}\label{Eq:Formula_1.31a}
	\end{equation}
	
	\noindent and the angular momentum is
	
	\begin{equation}
	H=\{GMa(1-e^{2})\}m\\  \label{Eq:Formula_1.31b}
	\end{equation}

	The model we shall use to simulate this situation is shown in Figure \ref{Fig:Satellite_three_masses}. The planet is represented by a point, $P$, and the satellite by a distribution of three masses, each $m/3$, at positions $S_{1},S_{2}$ and $S_{3}$, forming an equilateral triangle when free of stress. The masses are connected, as shown, by springs, each of unstressed length $l$ and the same spring constant, $k$. Thus a spring constantly stretched to a length $l'$ will exert an inward force

	\begin{equation}
	F=k(l'-l) \label{Eq:Formula_1.3.1}
	\end{equation}
	
	Now, we also introduce a dissipative element in our system by making the force dependent on the rate of expansion or contraction of the spring, giving the following force law:
	
	\begin{equation}
	F=k(l'-l)-c\frac{dl'}{dt}\label{Eq:Formula_1.32}
	\end{equation}
	
	\noindent where the force acts inwards at the two ends. It is the second term in Equation \ref{Eq:Formula_1.32} which gives the simulation of the hysteresis losses in the satellite \cite[18-19]{Woolfson1999a}.
	
\end{quotation}

\normalsize

Thus understood, the mathematical model includes classical Newtonian mechanics for describing all two-body systems under tidal stress, whether these are satellites, planets, or some other body. 

In order for this mathematical model to become a computer simulation, it must first be transformed into a \textit{simulation model}. As mentioned, such transformations are done by using formal methods (e.g., some discretization methods). For instance, equation \ref{Eq:Formula_1.31a} is described in the simulation model by the subroutine  $TOTM=CM(1)+CM(2)+CM(3)+CM(4)$; $EN=-G*TOTM/R+0.5*V2$, whereas the force equations represented in (\ref{Eq:Formula_1.31b}) and (\ref{Eq:Formula_1.3.1}) are described by subroutine ACC in \cite{Woolfson1999}. Besides these formal transformations, non-formal techniques purposely modify the mathematical model as well. As Figure \ref{Fig:Satellite_three_masses} shows, the mass of the satellite is not longer represented by one variable but rather three, each connected by springs of certain length. This is a good example for the claim that a simulation model retains parts of the structure of the mathematical model while it drops and adds new structures.\footnote{In fact, a simulation model includes a host of non-mathematical structures, such as loops, conditionals, subroutines, and other terminology that conceptually distance the simulation models from mathematical models.}

It is also important to notice that such simulation models represent a host of real-world phenomena, and only by means of \textit{instantiating} the set of initial and boundary conditions can researchers single out one specific satellite, orbiting around a specific planet, with a specific tidal stress, energy, and so forth. For instance, Woolfson and Pert use the following set of parameter values \cite[20]{Woolfson1999a}: \label{Parameter_values}

\small
\begin{quotation}
	number\_of\_bodies = $4$
	
	mass\_of\_planet = $2$ x $10^{27}$ kg
	
	mass\_of\_satellite = $3$ x $10^{22}$ kg
	
	initial\_time\_step = $10$ s
	
	total\_simulation\_time = $125000$ s
	
	body\_chosen\_as\_origin = $1$
	
	tolerance = $100$ m
	
	initial\_distance\_of\_satellite = $1$ x $10^{8}$ m
	
	unstretched\_length\_of\_spring = $1$ x $10^{6}$ m
	
	initial\_eccentricity = $0.6$
	
\end{quotation}

\normalsize

Thus instantiated, the computer simulation produces results that represent the behavior of a real satellite of a mass close to Triton, the largest moon of Neptune, orbiting around a planet with a mass close to Jupiter's. 

Thus understood, this simulation has considerable explanatory value. To see this, let me make the explanatory reasoning explicit. As an initial condition, the position of the satellite is at its furthest distance from the planet, hence the spikes only occur when they are at their closest. When this happens, the satellite is stretched by the tidal force exerted by the planet. Correspondingly, inertia makes the satellite tidal bulge lag behind the radius vector. The lag and lead in the tidal bulge of the satellite give spin angular momentum on approach, and subtract it on recession. When receding from the near point, the tidal bulge is ahead of the radius vector and the effect is therefore reversed. The spikes therefore occur because there is an exchange between spin and orbital angular momentum around closest approach \cite[21]{Woolfson1999a}. 

At this point, someone could bring forward Weirich's and Krohs' explanatory relation, and claim that we could obtain an explanation of the spikes by using the mathematical model. As discussed earlier, such an objection rests on the assumption that mathematical models are implemented as a computer simulation \textit{simpliciter}, taking the latter merely as an instrument of computation and ignoring the fact that the results have been modified by the computation of the simulation model. As the example of the simulation of a satellite shows, simulation models are conceptually and epistemically different from mathematical models, and only the former can be used to account for the explanation of the spikes.

This last point can be grounded on the fact that the results of the simulation show the patterns described in the simulation model. Take for instance the roundoff errors produced by the Runge-Kutta algorithm. Such errors are partly responsible for the orbital eccentricity trending steadily downwards, as shown in Figure \ref{Fig:Spikes}. We would be unable to explain this effect with the mathematical model alone, because it is the NBODY subroutine, responsible for implementing the Runge-Kutta method, that accounts for the orbital eccentricity. In this respect, the mathematical model by itself has limited explanatory force of the results of the simulation. This is the mistake that Woolfson and Pert make when they try to explain the spikes, as their explanation remains heavily dependent upon the mathematical machinery of the model and cannot account for an orbital eccentricity trending downwards \cite[21]{Woolfson1999a}. These facts provide a justification for taking the simulation model to be a complete representative of both the target system and the computation that represents the target system, and therefore as the most relevant unit with explanatory force. With these ideas in mind, let us now turn to explaining why the spikes in the simulation occur by making use of the simulation model.
	
\section{Explanation for computer simulations \label{Explaining_SP}}

Before discussing in detail the explanatory relation for computer simulations, I would like to provide two reasons why I believe that the unificationist account provides a suitable explanatory framework for computer simulations. First, the unificationist is a nomothetic theory of explanation, and as such the relation between explanans and explanandum depends entirely on our body of belief. Computer simulations are well suited for accommodation within this framework, since they are designed and coded using our current body of scientific knowledge. This contrasts with other theories of explanation, particularly ontic theories, where the explanatory relation depends on an objective external relation (i.e., causal relations \cite{Salmon1984a} and some sort of representation of causal relations \cite{Woodward2003}). This latter view is the one adopted by Weirich and Krohs, who consider the mechanicist as the most suitable framework for computer simulations. Admittedly, a discussion on this point is still necessary but for my present purposes the unificationist provides the right conceptual tools.
	
	\def\CSasUnification{This is not to say that computer simulations are unificatory systems. For such a claim we also need to specify in what respects they unify. Rendering a host of simulated phenomena -- some of which are clearly unknown -- is a core feature of computer simulations that squares well with the unificationist. A future task is to show in what respects there is unification in computer simulations, including models that are not straightforwardly unificatory.}
	
Second, unification consists of using the same patterns of derivation over and over, reducing in this way a multiplicity of phenomena that we have to accept as independent (i.e., phenomena for which we have no explanation, but for which one is nonetheless anticipated \cite{Barnes1994}). This is a core idea in the unificationist account that is also echoed by computer simulations. Computer simulations can produce a multiplicity of results of different kinds simply by setting their parameters to different values.\footnote{\CSasUnification} Variables and subroutines can take different forms: a mass can be a charge and instead of the force law the simulation could use the inverse-square law. This is something that Woolfson and Pert make clear in their analysis of the simulation, as well as in their code (see \cite{Woolfson1999}). Moreover, the same computer simulation highlights different aspects of the same target system. For instance, we might know why the satellite orbits around the planet, but have no idea why the spikes occur. The point here is that the occurrence of one set of results has no bearing on the likelihood that the next set of results will be known. 
	
Producing a multiplicity of results is not an \textit{ad hoc} characterization of computer simulations, but rather an inherent feature of these methods. It is then desirable for a theory of explanation to be able to account for, and capitalize on, these features. The unificationist is, in this respect, a suitable theory of explanation for computer simulations.

	\subsection{The unificationist framework} \label{The_unificationist_framework}

	Explanation, for the unificationist, begins with the set of accepted scientific beliefs, \textit{K}. In the sciences, \textit{K} can be interpreted as classical mechanics in physics, the evolutionary theory in biology, and the atomic theory in chemistry. Finding the set \textit{K} of accepted beliefs for computer simulations is in no way different from other areas of science, as the simulation model also relies on our current scientific knowledge. Examples from molecular biology can be drawn, as we simulate the effects of alanine scanning and ligand modifications based on molecular dynamics \cite{Boukharta2014}. Simulations in nuclear physics would include the Boltzmann-Uehling-Uhlenbeck model, some theorems from statistical mechanics such as Liouville's theorem, and general Hamiltonian equations of classical mechanics \cite{Hartmann1996}. And, of course, Woolfson and Pert's example relies on a set of differential equations as described by classical mechanics. 
	
	The real challenge for the unificationist is to specify what counts as the \textit{explanatory store over K, E(K)}. This is, what is the set of acceptable argument patterns that have explanatory force. According to  Kitcher, \emph{E(K)} encompass three main elements, namely, the \emph{schematic sentences} (i.e., expressions obtained by replacing some of the non-logical expressions in a sentence by dummy letters), a set of \emph{filling instructions} (i.e., the set of directions for replacing those dummy letters), and a \emph{classification} (i.e., the set of sentences that provide directions for which terms are to be regarded as premises, what is inferred from what, and so forth). The \textit{general argument pattern} (or \textit{argument pattern} for short) is ``a triple consisting of a schematic argument, a set of sets of filling instructions, one for each term of the schematic argument, and a classification for the schematic argument'' \cite[432]{Kitcher1989}. Additionally, there is the \emph{comments section}, that Kitcher uses for supplying additional information, such as minor details on the limits of an argument pattern, and possible corrections for it. To explain, then, consists of deriving descriptions of a multiplicity of phenomena using as few and as stringent argument patterns as possible. 
	
	In the context of computer simulations, the explanatory store is construed in similar ways. The \textit{schematic sentences} constitute expressions in the language of the simulation model. For instance, the schematic sentence would specify the subroutines NBODY, STORE, and ACC, variables like the mass of the different objects involved in the simulation, and diverse control flow statement (such as loops, conditionals, and the like). The \textit{filling instructions} makes use of the initial and boundary conditions set out earlier. As for the \textit{classification}, I take it to be just what Kitcher intended it to be, that is, the set of sentences that provide directions for which terms are to be regarded as premises, what is inferred from what, and so forth. Finally, the \textit{comments} section plays a more visible role than originally ascribed by Kitcher, since it works as a \textit{repository} for all the remaining information that have explanatory force, but that cannot be constructed as a schematic sentence. As I discuss in the next section an example of this is, again, the roundoff error responsible for the spikes steady trending downwards.

	\subsection{Example of an explanation} \label{Explanatory_relation}
	
	Allow me now to reconstruct the explanatory store for the simulation of an orbiting satellite under tidal stress. Following the unificationist, a possible explanatory schemata for explaining the spikes shown on Figure \ref{Fig:Spikes} are:\footnote{A reconstruction of computer simulations as arguments is given by \cite{Beisbart2012}. Unlike him, I am not claiming that \textit{all} computer simulations are arguments, but rather that some aspects of the simulation model can be reconstructed for explanatory purposes.}
	
	\begin{quotation}
		\noindent \begin{flushleft}
			\emph{Schematic Sentences}:\footnote{A full description of the variables, data types, and subroutines can be found in \cite{Woolfson1999}.}
			
			\par\end{flushleft}
		
		\begin{enumerate}
			\item There are two objects, $I$ and $J$, one of mass $CM(I)$ and another with a mass of $CM(J)$ $(\ll CM(I))$
			\item There is an orbit of semi-major axis $A$ and eccentricity $E$
			\item The object of mass $J$ is distributed into three masses, each $J/3$, at positions $POS(1)$, $POS(2)$ and $POS(3)$, forming an equilateral triangle free of stress.
			\item The relative velocities of the bodies are $VEL(1)$, $VEL(2)$, and $VEL(3)$
			\item \dots
			\item Subroutine \textit{NBODY} includes the Runge-Kutta subroutine with automatic step control.
			
			\item Subroutine \textit{STORE} stores intermediate coordinates and velocity components as the computation progresses.
			\begin{enumerate}
				\item For each position \{POS(1, 2, 3)\} of the satellite,  the expected orbital distance is given by equation:\\ $R=SQRT(POS(1)**2+POS(2)**2+POS(3)**2)$ 
				\item For the mass of the satellite \{$C(1, 2, 3)$\}, and mass of the planet $C(4)$, the expected intrinsic energy is given by the equation: \\
				Velocity $V(2)$ obtained from the square of relative velocities of the bodies,\\
				$TOTM=CM(1)+CM(2)+CM(3)+CM(4)$, and\\ 	$ EN=-G*TOTM/R+0.5*V2$
				\item For each position \{POS(1, 2, 3)\} and velocities \{VEL(1, 2, 3)\} of the satellite,  the expected intrinsic angular momentum is given by: \\
				\begin{minipage}[t]{10cm}
					$D1=POS(2)*VEL(3)-POS(3)*VEL(2)$,\\  $D2=POS(3)*VEL(1)-POS(1)*VEL(3)$, \\
					$D3=POS(1)*VEL(2)-POS(2)*VEL(1)$, and \\
					$H2=D1**2+D2**2+D3**2$
				\end{minipage} \vspace{1mm}						 
				\item \dots
			\end{enumerate} 
			
			\item Subroutine \textit{ACC} calculates the acceleration of each body due to its interactions with all other bodies
			\begin{enumerate}
				\item For each $I=2..NB-1$, $NB$ being the number of bodies, and for each $J=I+1..NB$, and for each $K=1..3$, equation $DIF(K)=XT(I,K)-XT(J,K)$ calculates the spring and relative forces, while the expected length of spring is given by the equation $ELP=SQRT(DIF(1)**2+DIF(2)**2+DIF(3)**2)$ 
				\item For each $I=1..NB-1$, $L=J+1..NB$, and $K=1..3$, the interactions for all pairs of bodies is given by (only the gravitational force is considered here): $R(K)=XT(J,K)-XT(L,K)$ and $RRR=(R(1)**2+R(2)**2+R(3)**2)**1.5$
				\item \dots\\
			\end{enumerate}
			\begin{description}
				\item [\textit{E}] \textit{The spikes formed are due to an exchange between spin and orbital angular momentum around closest approach.}\\
			\end{description}
		\end{enumerate}
		
	\noindent \emph{Filling Instructions}:
		
		\noindent The gravitational constant is set to $G=6.667E-11$. The mass $CM(I)$ and the mass $CM(J)$ will be replaced by a planet's and a satellite's respectively. A dissipative element is introduced into the structure by making the force dependent on the rate of expansion or contraction of the spring, giving a force law, where the force acts inwards at the two ends. Values for $POS(1)$, $POS(2)$, $POS(3)$, $VEL(1)$ and $VEL(2)$ must also be given.  Recall from earlier the parameter values set. \\

		\noindent \emph{Classification}:

		\noindent The classification of the argument indicates that 1-5 are premises, that 6, 7, and 8 are subroutines containing equations which can be obtained by substituting identicals. The explanandum \textit{E} follows from 6, 7, and 8 by derivation. \\
		
		\noindent \emph{Comments}:
		
		\begin{enumerate}
			\item Normal SI Units are used. 
			\item By changing the subroutines different problems may be solved. The CM's can be masses or charges or be made equal to  unity while the force law can be inverse-square or anything else (e.g., Lennard-Jones). \label{comments}
			\item The four-step Runge-Kutta algorithm is used. The results of two STEPS with TIMESTEP $H$ are checked against taking one STEP with TIMESTEP $2*H$. If the difference is within the TOLERANCE, then the two STEPS, each of $H$, are accepted and the STEPLENGTH is doubled for the next STEP. However, if the TOLERANCE is not satisfied, then the STEP is not accepted and one tries again with a halved STEPLENGTH. It is advisable, but not essential, to start with a reasonable STEPLENGTH; the program quickly finds a suitable value.
			\item The user is required to specify a TOLERANCE, the maximum absolute error that can be tolerated in any positional coordinate. If this is set too low, the program can become very slow. 
			\item Three kinds of forces are operating (subroutine $ACC$): the normal gravitational inverse-square law between all pairs of bodies, a second force due to the elasticity of the material as modeled by the set of three springs, and a third force operating between the three component bodies of the satellite and which depends on the rate at which the springs expand or contract. The third force provides the dissipation in the system.
			\item Subroutine \textit{NBODY} includes the Runge-Kutta subroutine with automatic step control, with a local discretization error on the order of $\mathcal{O}(h^p+1)$, and a total accumulated error on the order of $nCh^{p+1} = C(\bar{x}-x_0)h^p$. The method also has a roundoff error on the order of $\mathcal{O}(h^2)$ \cite[67]{Atkinson2009}.
		\end{enumerate}
		
	\end{quotation}
	
	Thus understood, the spikes of Figure \ref{Fig:Spikes} are explained as an exchange between spin and orbital angular momentum around closest approach. The explanation is obtained by derivation from the schematic sentences 1-8, as set forth in the filling instructions and the classification. 

	The \textit{comments} section, on the other hand, is a \textit{repository} of information that can be later used for explanatory purposes. As such, it plays two central roles. First, it documents boundaries and alternative explanations for the argument pattern (along with the limits and possible corrections). For instance, comment 2 indicates that the same explanatory schemata could be used for explaining masses as well as charges. This section also establishes standards and limitations. For instance, it is recommended to start with a reasonable STEPLENGTH, otherwise it could take longer to find suitable values for the Runge-Kutta algorithm. Similarly, an acceptable value for TOLERANCE is around 100 m. 
	
	Second, and more actively, it includes information about the results with relevant explanatory force, which could not be included as a schematic sentence. Reasons for this vary from formulations that play no role in the computing process (e.g., preconditions and postconditions), but that are essential in the assessment of the correctness of subroutines, to quantifications that are intrinsically difficult to reconstruct. For instance, the Runge-Kutta subroutine has a local discretization error function of the order $\mathcal{O}(h^p+1)$, and a total accumulated error on the order of $nCh^{p+1} = C(\bar{x}-x_0)h^p$ . Both are iterative functions, therefore the derivation of the local and total error depends on each iteration. Thus described, the local and total errors cannot be derived in the unificationist's preferred way, although both carry significant explanatory force. This situation allows us to expand the argument pattern and vouch for the comment section as a repository of relevant information for the explanation. 
	
In some simulations, it is possible to actually measure some of the errors with certain accuracy. To illustrate this point, suppose that the orbital eccentricity shown in Figure \ref{Fig:Spikes} is because of a small roundoff error such that, for each loop in the computation, a difference of $1^{-1000}$ kilometers is introduced for each revolution with respect to the real value. Although very small, this roundoff error plays a crucial role in the overall eccentricity trending steadily downwards. In particular, given a sufficiently large number of loops, such an error is responsible for the satellite reaching an eccentricity equal to 0 (i.e., after a determined number of runs, the satellite reaches a circular orbit). 
	
	In such a case, the errors are measured and reconstructed in terms of schematic sentences, allowing the derivation of the explanandum in the usual way. A possible reconstruction looks like the following:
	
	\begin{quotation}
		\noindent \emph{Schematic sentence: }
		\noindent 8') There is a discretization error of approx. $1^{-1000}$ in the \emph{total\_simulation\_time}. 
	\end{quotation}
	
	Reconstructing errors in this way is obviously the best option, as it allows the derivation of the explanandum. Unfortunately, measuring errors is not always an easy enterprise. For cases where errors are unmeasurable, like in the case of the Runge-Kutta above, I suggest to include them in the \emph{comments} section as further non-derivable reliable information with explanatory force. This move is perfectly acceptable within the unificationist framework, since non-derivable but explanatory information can be included in the comments section (Kitcher himself makes use of this resouce in section 4.6 \cite{Kitcher1989}). This is obviously not an ideal situation, as it restricts the explanatory force of the simulation. However, knowing about the presence of errors makes the epistemic difference between being aware of the existence of a disturbing factor, and thus being able to interpret and to explain the results of the simulation in light of those errors, and being unable to account for unexpected results. 

\subsection{Understanding} \label{understanding}
	
Thus far, my efforts have been directed towards showing the structure of the explanatory relation for computer simulations. This is the core issue behind a logic of scientific explanation for computer simulations and the main outcome of this article. Having said that, the logic of scientific explanation also has the larger aim of showing the epistemic gain of explaining. In the following, I give sense to the idea of understanding in the context of explanation for computer simulations.

To the unificationist, understanding comes from seeing connections and common patterns in what initially appeared to be brute or independent facts.\footnote{For an analysis of `brute' and `independent' facts, see \cite{Barnes1994} and \cite{Fahrbach2005}} `Seeing' here is taken as the cognitive maneuver of reducing the explanandum to a greater theoretical framework, such as our corpus of scientific beliefs. Schurz and Lambert take that, ``to understand a phenomenon \textit{P} is to know how \textit{P} fits into one's background knowledge'' \cite[66]{Schurz1994}. And Elgin asserts that ``understanding is primarily a cognitive relation to a fairly comprehensive, coherent body of information'' \cite[35]{Elgin2007}.\footnote{\cite{Barnes1992a} has rightly criticized theories of scientific explanation for failing to provide a full fledged account of what understanding consists in, and of how it is produced by scientific explanations. Here, I am only concerned with showing how understanding of results of computer simulations is realized within the unificationist account. It is therefore not within my interests to fully flesh out how such understanding is carried out. To this end, however, the `contextual approach to scientific understanding' as elaborated by \cite{Regt2005}, and the recent work on `how-possible' and `how-actually' understanding by \cite{Reutlinger2017} could shed some light on the issue.} Thus understood, such a theoretical reduction comes with several epistemological advantages. The explanandum becomes more transparent to us, we obtain a more unified picture of nature, we strengthen and systematize our corpus of scientific beliefs, and overall the world becomes a more simplified place (see \cite{Friedman1974}, \cite{Kitcher1981a, Kitcher1989}).


When results of a computer simulation are understood, a similar cognitive maneuver is performed: researchers are able to incorporate the results into a larger theoretical framework, reducing in this way the number of independent results seeking an explanation. Let us note that, for the case of computer simulations, such incorporation of the results is carried out in two steps: first, the results are included into the body of scientific beliefs related to the simulation model; and second, to include them into our greater body of scientific beliefs. By doing things in this way, we are fully within the unificationist framework. 
	
Let us illustrate this last point with the example of the results of simulating a satellite orbiting around a planet from section \ref{Example}. By explaining why the spikes occur (i.e., the results of the simulation shown in Figure \ref{Fig:Spikes}), we give reasons for their formation. Such explanation is possible, as discussed earlier, because there is a well-defined pattern structure that enables us to derive a description of the spikes from the simulation model. Understanding these results, then, comes from incorporating them into the larger corpus of scientific beliefs that is the simulation model. That is, researchers grasp how the results fit into, contributes to, and are justified by reference to the theoretical framework postulated by the simulation model. This is precisely the reason why we are able to explain the occurrence of the spikes as well as their downwards trend: both can be theoretically reduced by the simulation model. Furthermore, since the simulation model is dependent on well-established scientific knowledge -- in this case Newtonian mechanics -- we are able to see the results of the simulation in a way that is now familiar to us, that is, unified with our general body of established scientific beliefs about two-body mechanics. 



Thus far the standard image of the unificationist applies to computer simulations. But I believe that we can extend this image by showing how understanding results also encompass a \textit{practical} dimension. From the perspective of simulation research, understanding results also involves grasping the technical difficulties behind coding more complex, faster, and more realistic simulations, interpreting verification and validation processes, and conveying information relevant for the internal mechanism of the simulation. In other words, understanding results also feeds back into the simulation model, helping to improve our computer simulations. For instance, by explaining and understanding the reasons why the spikes trend downwards, researchers are aware of the existence of and have the means to solve roundoff errors, discretization error, grid resolution, etc. In computer simulation studies, researchers want to explain because they also want to understand and improve their simulations.

My last point is related to understanding real-world phenomena. As we discussed at the beginning, Weirich and Krohs had as their chief purpose the explanation of real-world phenomena. Both authors are right in thinking that the use of computer simulations is justified, in a large number of cases, because they provide understanding of certain aspects of the world. The question now is, then, can we understand real-word phenomena by explaining results of computer simulations?

I believe that we can answer these questions positively. We know that the visualization of the results of the simulation show two facets: first the spikes, which represent the behavior of a real-world satellite under tidal stress, and second their trend steadily downwards, which stands for a roundoff error coded in the simulation model. This means that the results of the simulation related to the spikes represent, and thus can be ascribed to, the behavior of a real-world satellite. In this sense, and following Elgin on this point \cite{Elgin2007, Elgin2009}, there is an entitlement -- via representation and ascription -- of the results of the simulation to the behavior of a real-world satellite. It is precisely because of this entitlement that we are able to relate our understanding of the results of the simulation with our understanding of the behavior of the real-world satellite. This point can also be made by means of the practical ability that presupposes understanding something. As Elgin cogently argues, the understander holds the ability to make use of the information at her disposal for practical purposes \cite[35]{Elgin2007}. In our case, researchers could actually build the satellite specified in the simulation.

On the other hand, the explanation of the spikes trending steadily downwards is carried out by factoring the roundoff errors in the explanatory store, as discussed in section \ref{Explanatory_relation}. Once derived, it becomes very clear why the results show spikes trending downwards, giving researchers a good idea of how to fix them. Now, since the trend does not represent and thus cannot be ascribed to an empirical counterpart, there is no entitlement. In this sense, our understanding remains confined to this facet of the results, and thus prevents us from wrongly claiming explanation of a circular orbit of the real-satellite. 


To contrast this last point, let us refer to Krohs once more. When he faces the question about computer simulations without a correlate in the real world, his only way out is to claim that simulations are of fictitious or imagined dynamic material system \cite[285]{Krohs2008}. Krohs then adds the burden of having to account for this fictional relationship, the way we explain it, and the way we understand it. Under my account, we do not need this fiction, but rather we can explain and understand results of computer simulations in their own right.

\section{Conclusions} \label{Conclusions}


This article presents and discusses a new approach to the logic of scientific explanation for computer simulations. I proposed to answer two questions, namely, `what is the explanatory relation for computer simulations?' and `what kind of epistemic gain should we expect?' The first question was discussed in section \ref{Explaining_SP} and the following subsections. The second question was answered in section \ref{understanding}. In order to address either question, it was first necessary to give reasons for taking the explanans as the simulation model and the explanandum as the results of the computer simulation. Reasons for this were given in section \ref{SE_and_CS}. 


The example of the simulation of the satellite used in section \ref{Example} gave us an intelligible reconstruction of a standard computer simulation, and therefore did not require us recreate it in full length. One possible reconstruction for explanatory purposes is shown in section \ref{Explanatory_relation}, but others could follow. For instance, simpler schematic sentences could be derived directly from the subroutines (i.e., NBODY, STORE, and ACC) provided that we know the input variables and the return value. Alternatively, we could rewrite some of the subroutines as more comprehensive schematic sentences. For instance, the subroutine that calculates the square of intrinsic angular momentum could be rewritten as: $SqIAM (P1, P2, P3, V1, V2, V3):D1,D2$, where \{P1, P2, P3, V1, V2, V3\} are the input variables and $D1$ and $D2$ are the return values. 

Additionally, more complex computational structures (e.g., libraries, data-bases, multi-scale computing, pseudorandom number generators for heuristic simulations, etc.) could be approached in a similar fashion, as long as the documentation is sufficiently informative of their functionality. In this sense, and regardless of how the computational structure is programmed, we can always know how it works, what sort of input variables are taken, and what values are  returned -- or should be returned. This moves the level of interpretation of the simulation model one step up, and solves the problem of limitations in reconstructing the simulation model as schematic sentences due to their complexity.

If the above considerations are correct, then my account accommodates explanation for computer simulations in a way that has no precedent in the literature. It then comes naturally that genuine philosophical issues would emerge which have not been addressed here. If this is the case, then more discussion on the topic should be expected.



\bibliography{/Users/Juan/Dropbox/00_PAPERS/BibTex/biblioZotero.bib}
\bibliographystyle{plain}

\end{document}